\documentstyle[12pt]{article}
\begin{document}
\begin{titlepage}
\begin{flushright}
Report \#RU 94-4B
\end{flushright}
\begin{center}
\LARGE {\bf Surface tension, hydrophobicity, and black holes:}
\end{center}
\begin{center}
\LARGE {\bf The entropic connection}
\end{center}

\vspace{3mm}
\begin{center}
{David J.E. Callaway*\\
Department of Physics\\
The Rockefeller University\\
1230 York Avenue\\
New York, NY 10021-6399\\
USA\\

callaway@physics.rockefeller.edu}
\end{center}

\vspace{3mm}
\begin{abstract}
The geometric entropy arising from
partitioning space in a fluid ``field theory''
is shown to be linearly proportional
to the area of an excluded region.
The coefficient of proportionality 
is related to
surface tension by a thermodynamic argument.
Good agreement with experimental data
is obtained for a number of fluids.
The calculation employs
a density matrix formalism developed previously
for studying the origin of black hole entropy.
This approach may lead
to a practical new technique for the evaluation
of thermodynamic quantities with important
entropic components.
\end{abstract}

\vspace{.40in}
\begin{center}
To be published in the Physical Review E.\\
PACS numbers:  68.10.Cr, 05.30.Ch, 61.20.Gy, 97.60.Lf
\end{center}

\vspace{.40in}

*Work supported by the U.S. Department of Energy
under Grant No. DOE91ER40651 Task B.

\end{titlepage}
\newpage
\hsize=6in.
\hoffset=-.5in

{\bf 1. Prolegomena}

\vspace{5mm}
The ground state of a quantum field theory
can be described by means of a density matrix.
If the field degrees of freedom inside a specified
volume are traced over, the result is a reduced
density matrix, $\rho_{out}$, which depends only
on the degrees of freedom external to the excluded
volume.  Associated with this reduced system is
a ``geometric entropy'', 
$S_{out}/k  = -  Tr (\rho_{out} \ln \rho_{out})$,
which quantifies the information lost by the partitioning
of space.  The geometric entropy is an expression of
the fact that $\rho_{out}$ lacks the information
contained in correlations between the excluded
interior volume and the rest of the system.

\vspace{5mm}
This density matrix formalism was recently utilized
in an attempt to provide a simple explanation of
the classic result$^{[1]}$ that the entropy
of a black hole is linearly proportional to
its surface area.
A spherical volume of a space containing
a free scalar field was excluded,
and the resultant geometric entropy was then determined.
It was found$^{[2,3]}$
that the geometric entropy of this 
``black hole'' is, in fact, proportional
to the area of the excluded region, rather than
to its volume.  This result seems initially to
be rather mysterious, since entropy (like
free energy) is generally
an extensive quantity.
However, in several important physical situations,
major contributions to the free energy of a
system are, in fact, proportional to its area.
These ``area law'' contributions are primarily responsible
for liquid surface tension and for ``hydrophobic'' effects
central to protein folding.

\vspace{5mm}
Although the density matrix formalism was developed
to further our understanding of black holes, it
does not directly utilize either classical general relativity
or quantum gravity.
It will be shown here that geometric entropy 
can be used to exhibit a point of commonality between
liquid surface tension and the aforementioned results,
which are germane to the study of black holes.
The geometric entropy arising
from partitioning space in 
an empirical fluid ``field
theory'' is calculated.
Remarkably, the geometric entropy is found
to be linearly proportional to the
area of the excluded volume, as in the 
above case$^{[2,3]}$ of a free scalar field.  A
thermodynamic argument relates the coefficient of
proportionality to liquid surface tension.
Good agreement with experiment is obtained for a number
of liquids, suggesting that the density-matrix formalism
may lead to a simple and accurate new way to evaluate
thermodynamic quantities with important entropic components.
Further anticipated developments
are mentioned in the closing remarks.

\vspace{5mm}

{\bf 2. A field theory of fluids}

\vspace{5mm}
The applicability of the density-matrix formalism to fluids
is founded upon the observation that a liquid at
finite temperature can be considered, for some purposes, as
the vacuum state of a field theory.
This is a natural step to take, since the structure
and properties of a fluid at equilibrium can be described by 
its molecular distributions$^{[4,5]}$.  These distributions can, in
turn, be used to define the Green's functions of a 
field theory, e.g.,

\begin{eqnarray}
\left <\rho ({\bf r'}) \right > & = & \rho_0 \nonumber\\
\left <\rho ({\bf r'})\rho ({\bf r'} + {\bf r}) \right > & = &
\rho^2_0[1 + h({\bf r})] + \rho_0 \delta({\bf r})
\end{eqnarray}
where the right-hand sides of Eqs. (1) are independent of
${\bf r'}$ for a homogeneous fluid.  These Green's functions
$\left <\cdots \right >$ are vacuum (i.e., ground state)
expectation values of fluid density fields
\{$\rho({\bf r})$\}, calculated by means of a functional integral
measure defined below.  The complete set of
Green's functions, which can be taken as the definition
of a field theory, can likewise be expressed in terms
of higher-order molecular distributions.
The average fluid number density $\rho_0$ and
pair correlation function $h({\bf r})$ are physically measurable,
and can be extracted from experimental data.  The first two
Green's functions are thus known quantities.

\vspace{5mm}
The idea of representing a fluid at finite temperature
by the vacuum state of a field theory is quite novel,
and therefore deserves careful explanation.
Relations such as Eqs. (1) are frequently found
in the classical treatments of fluids$^{[4,5]}$.
However, in the ``classical'' treatment,
the averages $\left <\cdots \right >$ are meant to be taken
over canonical or grand canonical ensembles.
Thus, although the definitions Eqs. (1) are 
identical to equations commonly found in textbooks,
their interpretation is quite different.

\vspace{5mm}
In the classical formalism, one usually begins by taking averages
over a canonical ensemble of N particles which interact
via a potential function $V({\bf r_1,r_2,\cdots ,r_N})$.
Then the canonical ensemble averages 
$\left <\cdots \right >_{canonical}$ of densities
and their correlations are defined by relations like

$$\left <\rho ({\bf r}) \right >_{canonical}  =  Z_N^{-1} \times
\int d{\bf r_1} \int d{\bf r_2} \cdots \int d{\bf r_N} 
\exp (-\beta V) \times \rho_{classical} ({\bf r})$$

where

$$\rho_{classical} ({\bf r}) \equiv
\sum_{i=1}^N \delta ({\bf r-r_i})$$
The integration is over the coordinates
of all particles 1 through N, $\beta \equiv 1/kT$ is the
inverse temperature, and $Z_N$ is the partition function.
Grand canonical averages can then be calculated in
the usual fashion to yield the familiar molecular 
distributions.

\vspace{5mm}
One can thus generate
an infinite set of grand canonical expectation
values $$\left <\rho ({\bf r_1}) \rho ({\bf r_2)} 
\cdots \rho ({\bf r_M}) \right >_{gc}$$
for all products of the quantity $\rho_{classical} ({\bf r})$.
It is then that a novel step is taken.  Each grand 
canonical average $\left <\cdots \right >_{gc}$ is identified with
the corresponding Green's function of a certain
field theory.  Because the complete set of grand canonical
expectation values is known in principle, 
all of the Green's functions of this field theory are
also known.  The full set of Green's functions for
a field theory can then be taken as a definition of that
field theory.
One important point must be emphasized, however:
field theoretic Green's functions are,
{\it by definition}, expectation values of certain operators
taken over the vacuum state.  Thus, if the operators
and field-theoretic ground state are chosen so as to reproduce
the hierarchy of grand-canonical expectation values
discussed above, 
the ground state of a field theory provides
a description of a fluid that is 
{\it precisely equivalent} to
knowledge of the complete set of molecular
distributions.

\vspace{5mm}
In practice, of course, neither field theories nor
grand-canonical fluid models are generally solvable.
Thus, one must resort to judicious approximation.
As the present study is primarily intended to illustrate
the potential benefits of a very new approach, 
the field theory used is a simple one, a generalized
free field theory chosen to reproduce 
only the first two Green's functions Eqs. (1).

\vspace{5mm}
The simplest field theory whose Green's functions reproduce 
Eqs. (1) has a ground-state density matrix proportional to
$\exp[-\frac{1}{2}(S\{\rho\} + S\{\rho'\})]$, where
S is quadratic in the field
variables \{$\rho({\bf r})\}$:

\begin{eqnarray}
\rho_0 S\{\rho\} &=& \frac{1}{2}\int d{\bf r}\int d{\bf r'}\;[\rho 
({\bf r}) - \rho_0] W({\bf r - r'})[\rho({\bf r'}) - \rho_0] \nonumber\\
~&~\nonumber\\
W({\bf r - r')} &\equiv&  \delta({\bf r-r'})
- \rho_0c({\bf r - r'})
\end{eqnarray}
and the function $c({\bf r - r')}$ remains to be
determined.  As an additional convenience, the field variables
\{$\rho({\bf r})$\} are allowed to range from negative
to positive infinity.  The expectation values in Eqs. (1) are
calculated via the functional integral

\begin{eqnarray}
\left <{\cal O} \right > & \equiv & \Xi^{-1} \int \cal{O} \{\rho\} 
\exp [-S \{\rho\}] \cal{D}\rho \nonumber \\
\Xi & \equiv & \int \exp[-S \{\rho\}] {\cal D}\rho
\end{eqnarray}
The function $c({\bf r})$ is then required via
Eq. (1) to satisfy

\begin{equation}
h({\bf{r}}) = c({\bf{r}}) + \rho_0 \int
c({\bf{r - r'}}) h({\bf{r'}}) d{\bf{r'}},
\end{equation}
which is the well-known Ornstein-Zernike relation$^{[6]}$
linking the direct correlation function $c({\bf r})$ with the
pair correlation $h({\bf r})$.  The physical meaning of 
the function $c({\bf r})$ is thus manifest.
The success of the Ornstein-Zernike approach derives in
part from the short-ranged character of $c({\bf r})$,
suggesting that Eq. (2) is, indeed, a sensible starting
point.

\vspace{5mm}
For this choice of c({\bf r}), Eqs. (2) and (3) 
reproduce Eqs. (1).  The
constraints posed by Eqs. (1) alone are, however, insufficient 
to define a field theory uniquely.  The generalized free
field theory implicit in Eqs. (2) and (3) is only the simplest
solution of Eqs. (1), and therefore provides an approximate
description of a real liquid.  In the theory defined here,
all higher-order correlations are calculable in terms of 
products of those in Eqs. (1).  
The fluid theory defined by Eqs. (2)
possesses the property that
higher-order molecular distributions factorize,
which is known$^{[5]}$ to be correct for large separations.
There is no barrier of principle
to performing a more elaborate calculation
with a better model of a fluid; however,
the labor of calculation would be larger.
(This point is discussed in further detail
below).
The integration over unphysical
negative values of the fluid density fields $\{\rho({\bf r})\}$ 
is also permissible, for $S\{\rho\}$ is
positive definite and is sharply peaked about 
$\rho({\bf r})\cong\rho_0 > 0$, provided that the Fourier transform
$\tilde{W}(k)$ is positive (as it is for a fluid
with finite compressibility).  Although 
negative, ``virtual'', values of the density appear formally
in the functional measure Eq. (3), negative densities do
not explicitly enter into the physically 
observable molecular distributions,
which are the Green's functions of the theory.
The standard$^{[5]}$ density functional approach
can be
recovered by adding the usual potential term
${\int v({\bf r}) \rho({\bf r}) d{\bf r}}$
 to $S\{\rho\}$,
and then evaluating $\Xi$ as a functional of the average
densities 
$\bar \rho({\bf r}) \equiv \left <\rho ({\bf r}) \right >_v$.
The density functional $\Omega\{\bar \rho\}$ of the bulk fluid is
then simply $\Omega\{\rho_0\}-kT \times \ln [\Xi\{\bar \rho\}]$,
which is
the ``effective action'' of the field
theory Eq. (3). For any $S\{\rho\}$, $\Omega\{\bar \rho\}$
calculated in this fashion is 
convex$^{[7]}$ for a uniform system
and can have no more than one
minimum, so no ``Maxwell construction'' is needed.

\vspace{5mm}
In what follows, the fluid density $\rho_0$ 
and pair correlation function $h({\bf r})$ are
taken from experimental results.  No attempt is made to
derive either of these quantities from first principles.
One final simplification is therefore 
employed in the sequel.  Experimental
techniques (such as x-ray scattering) typically measure only
a spherical average $h(|{\bf r}|)$ of the pair correlation.
Although the above formalism can be easily applied to test a
nonspherically symmetric {\em model}, using real data
requires the assumption that $h({\bf r}) = h(|{\bf r}|)$ and
$c({\bf r})$ = $c(|{\bf r}|)$.  This assumption is, however,
commonplace$^{[12]}$.

\vspace{5mm}

{\bf 3. Relating geometric entropy to surface tension}

\vspace{5mm}

The geometric entropy can be related to fluid
surface tension by
a simple argument.
A well-known theorem of statistical
mechanics$^{[8]}$ states that the probability of
observing a fluctuation in a system at 
temperature T is given by
$\exp (-W/kT)$, where W is the reversible work required
to produce the fluctuated configuration through the
application of a constraint, and k is Boltzmann's 
constant.
Thus the reversible work required to produce
a cavity of low density in a fluid can be found from 
the probability that such a cavity occurs
via a fluctuation.  
Consider a subsystem of the fluid contained
within an imaginary sphere of radius $R$.
The density matrix $\rho_{in}$ for this
subsystem can be constructed from that given above for
the full volume by integrating out the field
variables \{$\rho({\bf r})$\} outside the sphere.
This density matrix is a weighted
sum of the states internal to
the sphere, given that the exterior region
is unobserved.  It is shown below that,
ignoring volume work,
the relative probability of occurrence
of any state which approximates
a low-density cavity is essentially
$\exp (-S_{in}/k )$, where
$S_{in}/k  = -Tr (\rho_{in} \ln \rho_{in})$
is equal to the geometric entropy $S_{out}/k$ of the subsystem.
The macroscopic surface tension $\gamma$ for a fluid 
at a fixed temperature $T$ is then
the large $R$ limit of $T S_{in} (4\pi R^2)^{-1}$.
Since the Green's functions used to specify the
fluid field theory depend implicitly upon the
liquid temperature, geometric entropy
surface tension
varies nonlinearly with temperature, and thus
contains contributions from
both surface excess entropy and enthalpy.
\vspace{5mm}

The useful correspondence between this surface tension
calculation and the aforementioned ``black hole''
entropy result arises upon construction of the
complementary density matrix $\rho_{out}$.
In this complementary case, only the fields
\{$\rho({\bf r})$\} {\em inside} the sphere are
integrated out to yield $\rho_{out}$.  Then
it can be shown$^{[3]}$ that
$S_{in}/k  = S_{out}/k  = -  Tr (\rho_{out} \ln \rho_{out})$,
where $S_{out}/k$ is the geometric entropy of the subsystem.
The density-matrix calculations$^{[2,3]}$
relevant to black hole entropy involve the development
of techniques for calculating $S_{out}$; by the reasoning
given here, they are also useful for extracting
liquid surface tension.
\vspace{5mm}

The density matrix calculation proceeds as follows.  When
the field variables \{$\rho({\bf r})$\} outside the sphere 
are integrated out, the result$^{[2]}$ is a reduced density matrix

\begin{equation}
\rho_{in} \{q,q'\}  =  [det(M/\pi)]^{\frac {1} {2}}
\exp [-\frac {1} {2}
(qMq+q'Mq')-\frac {1} {4} (q-q')N(q-q')]
\end{equation}
where \{$q({\bf r}) \equiv [\rho({\bf r})-\rho_0]$\} inside the sphere.
The matrix $M$ is the inverse of $W/(2\rho_0)$,
taken over this interior domain, while $N \equiv W/(2\rho_0)-M$.
Since this domain is finite,
$M$ and $N$ are to be considered sums over discrete values
of wavenumber.
\vspace{5mm}

The reduced density matrix $\rho_{in}$ is next expressed in a new
basis \{$x$\} in which the matrix 
$\Lambda_S \equiv M^{-{\frac {1} {2}}}NM^{-{\frac {1} {2}}}$
is diagonal,
with eigenvalues $\lambda$.  
The
coordinates \{$x$\} are normalized by $x^2 = qMq$, so that the
determinant prefactor in Eq. (5) is eliminated.
Then$^{[2,3]}$ for each mode [with eigenvalue
$\lambda \equiv 4\xi (1-\xi )^{-2}, 0 \le \xi \le 1$]:

\begin{equation}
\rho_{in, \lambda} \{x,x'\}  = \sum_{n=0}^\infty p_n \Psi_n (x) \Psi_n (x')
\end{equation}
\begin{equation}
\Psi_n (x) \equiv (2^n n! \times \root \of {\pi / \alpha })^{-\frac {1} {2}}
H_n(\root \of {\alpha} x) \exp (-\alpha x^2/2)
\end{equation}
where $H_n(z)$ is an Hermite polynomial,
$\alpha \equiv (1+\xi )/(1-\xi )$, and $p_n = (1-\xi )\xi^n$.
The reduced density matrix is a product of those for each
mode $\lambda$:
\begin{equation}
\rho_{in} \{x,x'\} = \prod_{\lambda} \rho_{in, \lambda} \{x,x'\}  
\end{equation}
Each mode $\rho_{in, \lambda}$ behaves like a thermal density matrix
for an harmonic oscillator, specified by a frequency $\alpha$ and
effective temperature $T_{eff} = \alpha /\ln(1/\xi)$.  
\vspace{5mm}

The probability that a mode $\lambda$ of the spherical
subsystem is found in state $n$ is thus $p_n$.  Associated with
this probability distribution are two useful quantities.  These
are $\bar n$, which is the average state number:

\begin{equation}
\bar n = \sum_{n=0}^\infty n \times p_n = \xi /(1-\xi )
\end{equation}
and $S_{in}$,
the ``geometric entropy'', in terms of which the density of
states is $\exp (-S_{in}/k)$:
\begin{equation}
S_{in}/k = -\sum_{n=0}^\infty p_n \ln p_n 
= -\ln (1-\xi) - \bar n \ln \xi
\end{equation}

The probability of
finding the mode in any state $n \ge N$ is $p_{any}(N)=\xi^N$.
Thus, for $n \ge \bar n$,
\begin{equation}
p_n \le p_{\bar n} = \exp(-S_{in}/k) \le p_{any}(\bar n)
\end{equation}
For the systems of interest here, $\xi$ is generally
quite small, so
the second inequality is nearly an equality:
\begin{equation}
\exp (-S_{in}/k) \cong p_{any}(\bar n)
\end{equation}

\vspace{5mm}

In the classical methods of calculation, the probability that
the system attains a given density is calculated.  The course
taken here is, however, conceptually quite different.
The quantity of interest here is the relative probability
to find the subsystem in certain {\em states}, those
whose characteristic density is relatively low.  For a
given state, there is a finite probability that any 
value of the 
density will be attained.  The expectation value of an
operator can, however, be determined for such a
state.  Thus, the expectation value of $x^2 \sim (\Omega \{ \rho
\} - \Omega \{ \rho_0\})/kT$ in state $n$ is given by
$(n+\frac {1} {2})/\alpha$.  The probability distribution 
$[\Psi_n (x)]^2$ for $n > 0$ typically is maximized when
$x \cong x_{peak} \equiv \pm [(2n+1)/\alpha ]^\frac {1} {2}$,
corresponding to values of the density that are above and
below the bulk value.  A state whose density is expected to differ
from the bulk value by at least the inverse volume of the
subsystem (so that at least one molecule is absent) must
therefore have $n \ge \bar n$.  By Eq. (12), the probability
of finding the system in such a state is well approximated
by $\exp (-S_{in}/k)$.
\vspace{5mm}

Where, then, are the approximations made,
and how does one deal with more complex situations?
Let us review the logic of the calculation.
One begins with a field theory in
its ground state.
The question of interest is the probability that
a given fluctuation occurs inside an imaginary sphere
of radius $R$, regardless of what occurs outside
that imaginary sphere.  Thus, the degrees of freedom
outside the imaginary sphere are integrated out.
What remains is a field theory that is defined in
terms of variables internal to the sphere.
This field theory gives the relative probability
of a given internal configuration, assuming
that the region external to the sphere is unobserved.
In this regard, there is a point of contrast between
the fluid calculation and the black hole entropy
result.  In the latter case, degrees of freedom
inside the black hole are considered {\it unobservable}
on fundamental grounds, rather than simply being
{\it unobserved}.  Nevertheless, the correct
formal procedure is the same in the two cases--the
unobserved variables are integrated out.
\vspace{5mm}

It is important to realize that, at least in principle,
no approximations need to be made to reach this
stage of the calculation.  
Instead
of the quadratic form given in Eq. (2),
one could just
as well begin with an arbitrarily complex
density functional $S\{ \rho \}$,
corresponding to an arbitrary set
of molecular distributions. 
The virtue of the simple form Eq. (2)
is that it allows the calculation of the reduced
density matrix to be made explicit.
Nevertheless, by perturbative or
numerical means, it is
presumably possible to evaluate
the density matrix for the system
inside the imaginary sphere for an arbitrary
density functional $S\{ \rho \}$.
The result for each eigenmode of the density
matrix will be formally identical to Eq. (6),
although the eigenfunctions $\Psi_n$ will no longer
be the explicit forms of Eq. (7).
Thus, the simple generalized free-field
theory implied by Eq. (2) is a calculational
convenience, rather than an integral part of
the formalism.  Moreover, as is seen below,
this simple functional form suffices to reproduce
not only the necessary area-law form of the
geometric entropy (implying a constant surface
tension), but gives reasonable answers for
the surface tension as well.
\vspace{5mm}

The next step in the analysis is the calculation of
the probability to find the interior region in
states whose expected density is less
than the bulk value.  
Another point must be made clear here.
Although the surrounding imaginary sphere represents
a sharp boundary, the interface itself is not constrained
to be as sharp.  It is only necessary that
the boundary of the imaginary sphere be
placed outside the interface region.
Thus, one could imagine making
the imaginary sphere extremely large, and asking
the probability to observe states of density
only slightly less than the bulk value.  For
a sufficiently large sphere and small deviation
from bulk density, the quadratic 
approximation to the density functional $S\{ \rho \}$
must therefore be valid.
For a small sphere,
there will however be an error in estimating the
size of the bubble of low density contained
within the sphere.
The size of the interface region
should be dictated by molecular dimensions.
(In other words, the existence of an interface
in a non-critical system should not
perturb the system strongly more than a few 
molecular diameters away from that interface).
Thus, the ratio of the interface width to
the sphere size should vanish for a large
enough sphere (which is, of course, the
result of interest for macroscopic 
surface tension).  Therefore, for a large
enough sphere, the area of the low-density bubble
is asymptotically equal to the area of the surrounding
imaginary sphere.
\vspace{5mm}

If a different problem is formulated,
such as the calculation of the ``boundary'' free
energy associated with a bulk liquid in contact
with a hard wall,
constraints on the selected configurations
must be introduced.  
For the boundary free energy problem, the density
must be zero at the wall.
Thus, rather than asking 
[as in Eq. (12)] the probability $p_{any}(N)$
to find a mode in any state greater than $N$,
one asks the probability to find states
with a node in the proper location.
A more detailed
analysis would be needed in order to select
the proper states to include in this sum.  
However, in any case, 
there must be factors of $p_n$ included in the calculation,
and thus there should also always be a connection to
the geometric entropy.
This relation should persist even if a better 
model of a fluid than Eq. (2)
is used.
\vspace{5mm}

At least for some problems,
it is possible to improve upon the
results given by the approximate quadratic
density functional $S\{ \rho \}$ by means
of a simple perturbative analysis, although
in general such improvements likely require
numerical simulation.
One example of how a perturbative
study might proceed is as follows.
Consider a more elaborate model of
a fluid constructed
by replacing
the probability functional
$\exp [-S\{\rho\}]$ defined by Eq. (2) with a new functional
$$\exp [-S_{new}\{\rho\}] = \exp [-S\{\rho\}] +
 \epsilon \times \exp [-S_{gas}\{\rho\}],$$
where $\rho_0 S_{gas}\{\rho\}=
{1 \over 2}{\int [\rho({\bf r})]^2 d{\bf r}}$.
Here, $\epsilon$ is a small number that gives the
relative probability of the
bulk fluid to be in the gas phase.
The new action, $S_{new}\{\rho\}$, has two minima,
essentially at zero density, $\rho=0$, and at the bulk density,
$\rho=\rho_0$, and thus includes
a crude characterization of the gas
phase of the fluid.
One can then calculate, using ordinary quantum-mechanical
perturbation theory, the change
in the eigenvalues $\varrho_n$ of
the density matrix $\rho_{in}$ due to the additional
term in $\exp [-S\{\rho\}]$.
The gas-phase contributions to the geometric entropy
$-\Sigma \varrho_n \ln \varrho_n$
can then be expanded in a
series in the
small number $\epsilon$, allowing
geometric entropy
to be determined in a controlled
fashion.
\vspace{5mm}

It is therefore clear that
no barrier of principle exists in
applying the techniques developed
here to fluid models of arbitrary complexity.
The present formalism is thus not subject
to the easy criticism$^{[9]}$ applied to early
integral-equation treatments of surface tension.
In these early calculations, the direct correlation
function appropriate to a bulk fluid was
employed essentially to estimate the relative
probability of various interfacial configurations,
whose local density was necessarily far from
its bulk value.  These calculations were thus
highly sensitive to the tails of the probability
distribution employed, while the bulk direct correlation
only provides information about the distribution
near its peak.  The present argument
obviates this difficulty by counting states,
since the specific form of the density matrix for
densities far from the bulk value contributes little
to $S_{in}/k  = -  Tr (\rho_{in} \ln \rho_{in})$.
Thus, by calculating the geometric entropy, one
is determining the relative probability of the
system to be in a state whose typical 
density is {\em outside} the region where the
probability distribution is maximized.
Since both the distribution near its peak
and the overall normalization of the probability
distribution are known, this calculation can be
done relatively accurately.
Moreover, as the density matrix $\rho_{in}$
is constructed by integrating over all fields \{$\rho({\bf r})$\}
external to the sphere, correlations of arbitrarily
long wavelength are included, as they should be$^{[5,9]}$.
Nevertheless, as discussed above,
if the effects of complex boundary conditions
are to be accounted for in a satisfactory way, further
constraints on the available states must be included.

\vspace{5mm}

{\bf 4. Evaluation of surface tension} 

\vspace{5mm}

The calculation of $S_{out}$ is easily performed using Eq. (34)
of ref. [2].  The asymmetric matrix 
$\Lambda \equiv M^{-1}N$ has the same eigenvalues as
the symmetric matrix
$\Lambda_S \equiv M^{-{\frac {1} {2}}}NM^{-{\frac {1} {2}}}$,
but is easier to construct numerically.  Therefore,
the eigenvalues of $\Lambda$ are computed.
By making an expansion in spherical harmonics, the
problem reduces to finding the eigenvalues $\lambda_{\ell}$ of
the matrix

\begin{equation}
\Lambda_{\ell} (r, r'') = - \int^R_0T_{\ell}[1/\tilde{W}]
(r,r') \;\; T_{\ell}[\tilde{W}](r',r'')\:\:r'^2dr'
\end{equation}
with respect to the measure $(r'')^2dr''$
over the interval
\{$r, r'' >R$\}, where

\begin{equation}
T_{\ell} [\tilde{W}](r',r'') \equiv (r'r'')^{-\frac{1}{2}} 
\int^{\infty}_0 k\,d\,k\, \tilde{W}(k)\,J_{\ell+{\frac{1}{2}}}(kr')
\; J_{\ell+{\frac{1}{2}}} (kr'')
\end{equation}
and similarly \,for \, $T_{\ell}[1/\tilde{W}]$.  Here, \,
$\tilde{W}(k)$ is the Fourier
transform of \, $W(r)$, \, while \\
$J_{\ell+{\frac{1}{2}}}$ $(z)$ is a
Bessel function.  Then$^{[2]}$

\begin{equation}
S_{out} /k= \sum^\infty_{\ell=0} (2 \ell+1) \{ \ln({\frac{1}{2}} \lambda
^{\frac{1}{2}}_{\ell}) + (1 + \lambda_{\ell})^{\frac{1}{2}} \ln
[(1+\lambda_{\ell}^{-1})^{\frac{1}{2}} + \lambda_{\ell}^{-\frac{1}{2}}]\}
\end{equation}

The sum in Eq. (15) is taken over all eigenvalues $\lambda_{\ell}$
for each $\ell$.
The function $\rho_0\tilde{h}(k) = 1/\tilde{W}(k) - 1$ is taken 
directly from experiment, obviating the need for an independent
determination of the density $\rho_0$.
As k increases, $\tilde{h}(k)$ goes
to zero.  Hence, the substitutions 
\begin{equation}
1/\tilde{W}(k) \rightarrow
\rho_0 \tilde{h}(k) \equiv \tilde w^{-1}(k) 
\end{equation}
\begin{equation}
\tilde{W}(k) \rightarrow -\rho_0
\tilde{h}(k)/[1+\rho_0\tilde{h}(k)] \equiv \tilde w(k) 
\end{equation}
were made in Eqs. (13-14)
to ensure the convergence
of the integrals Eq. (14).
Since $T_{\ell}[1](r,r') = 
(r)^{-2} \delta(r - r')$, subtracting one from
$\tilde{W}(k)$ [or $1/\tilde{W}(k)]$ leaves $\Lambda_{\ell}(r,r'')$
unaffected for $\{r,r''>R\}$, and so the sum
Eq. (15) is unchanged by this replacement.
In contrast to
the case$^{[2,3]}$ of a massless
scalar field theory [where $\tilde{W}(k)=k$],
no ultraviolet cutoff is needed here, since
$\tilde{h}(k)$ vanishes for large k.
Data are generally presented as a
set of $N$ values of $\rho_0\tilde{h}(k)$ at uniformly spaced
intervals of $k=n \times (\Delta k)$,
so the integrals Eq. (14) can be viewed as
finite trapezoid-rule sums.  The eigenvectors of $\Lambda _{\ell}
(r, r'')$ are linear combinations of 
\begin{equation}
u_{\ell}(k,r'')=\Theta
(r''-R)(r'')^{-\frac{1}{2}} J_{\ell+\frac{1}{2}}(kr''),
\end{equation}
for
values of $k$ contained within the data set,
reducing the problem
to a numerical diagonalization of the
$N \times N$ matrix 
\begin{equation}
\Lambda_{\ell}(m,n)= -m [\tilde w^{-1}(m \Delta k)]
\sum_{i=1}^N I_{\ell}(m,i) [\tilde w(i \Delta k)]
[\delta_{in} 
- i I_{\ell }(i,n)]
\end{equation}
for each $\ell $.  The required 
indefinite integrals of Bessel functions
\begin{equation}
I_{\ell }(m,n) \equiv \int_0^{R \Delta k} 
J_{\ell+\frac{1}{2}} (ms)
J_{\ell+\frac{1}{2}} (ns) \: s \: ds
\end{equation}
can be performed
analytically. 
The computational 
procedure was checked against the test case 
$h(r) = -\exp (-mr)$,
for which the integrals Eq. (14) are known 
exactly.

\vspace{5mm}

The geometric entropy $S_{out}(R)$ for water at 25$^{\circ}C$ was
calculated by applying the above procedure to structure
function data$^{[10]}$.  The summand of Eq. (15) vanishes
exponentially with $\ell$, albeit with a large decay length
$(\sim 10)$.  By including all values of $\ell \leq 75$, $S_{out}
(R)$ can be accurately calculated up to $R\sim7.5\AA$.  
Figure 1 displays the result.
Except for
very small R values, $S_{out}(R)$ is
linearly proportional to $R^2$, as it
is for a cutoff scalar field
theory$^{[2,3]}$.  Asymptotically, $S_{out}(R)
\sim 2.50 R^2$, yielding a surface tension of 82 dynes/cm, 
which is 14\% larger than the experimental number 72 dynes/cm
$^{[11]}$.  
For comparison, scaled particle theory using the
usual hard-sphere radius of $2.7\AA$ predicts$^{[13]}$ 52
dynes/cm at $303^{\circ}K$.  A detailed molecular dynamics
simulation$^{[14]}$ of water at 305$^{\circ}K$ yields 67 
dynes/cm, to be compared with the 71 dynes/cm obtained 
experimentally at this temperature.  These simulation 
results can be {\em fitted} by scaled particle theory, 
if the significantly larger value of $2.875 \AA$ is chosen
for the hard-sphere radius of water.
[As there are several
implicit scales present, dimensional analysis cannot
be used to obviate the density-matrix entropy calculation.
Scaled particle theory, where surface tension is
given in terms of the hard-sphere radius $a$ by
the form $\gamma = kT/a^2 f(\rho_0 a^3)$, is probably
the simplest viable alternative.  Indeed,
the only external input to the density matrix
calculation, $\rho_0\tilde{h}(k)$,
is dimensionless.]
\vspace{5mm}

The surface tension of a number of other liquids was
also calculated using their 
measured structure functions$^{[15,16]}$.
The density matrix results
(DM) are given in Table 1, along with experimental data 
$^{[11,13,17]}$ and available estimates$^{[13]}$ from scaled particle
theory (SPT).
The agreement between the density matrix
results and experimental data over a wide temperature
range is respectable, especially since the only input to
the calculation is the (measured) structure factor.  By
contrast, scaled particle theory is quite sensitive to
the value chosen for the hard sphere radius.  This
parameter is not measurable directly, and is usually
determined by fitting compressibility data.
The closest agreement between experiment and the density matrix
calculation occurs for sodium, potassium, and water, whose 
structure functions are known with the best precision.

\vspace{5mm}

{\bf 5. Conclusions}

\vspace{5mm}

It is worth emphasizing the remarkable fact that the geometric
entropy was found to be essentially a linear function of
the area in every case considered here, in addition to
the free scalar field theory considered previously by
others $^{[2,3]}$.  On very general grounds$^{[3]}$, the eigenvalues
of the reduced density matrix (and thus the geometric 
entropy) depend upon the area of the 
boundary surface, and not upon the volume of the excluded
region.  However, it is not clear why the geometric entropy
should be linearly proportional to this area, especially
for situations whose physical basis and mathematical
structure are substantially different.
(Thus, for example, the calculation of geometric
entropy in the ``black hole'' case required the
introduction of an ultraviolet cutoff; here, since
the structure factor of a fluid vanishes at large
wavenumber, no cutoff is needed.)
Moreover, the linear dependence of geometric
entropy on area provides
an additional point of support
for the present mode of calculation.
It not only gives the correct ``area law'' form for
this contribution to the free energy, but gives
reasonable values for the constant of proportionality (which is
the surface tension).
\vspace{5mm}

As a rule, few ideas relevant to quantum gravity have practical
applications.  One exception was given here.  By determining
the geometric entropy of a ``black hole'' 
in a fluid ``field theory'',
the liquid surface tension can be evaluated in a novel way.  
These calculations are essentially exact results for 
a field theory that is an approximation to a real liquid.  The 
procedure is therefore automatically  self-consistent, which 
may partly account for its plausible accuracy.  Possible
extensions of the formalism include an extraction of the
``hydrophobic'' potential of mean force, which is a major
ingredient of protein folding models$^{[18]}$.  This may be
determined by excluding two regions of fluid and calculating
the resultant geometric entropy as a function of the distance between
them.  The density matrix method is sensitive to the form
of the structure factor over its entire range.  Thus, the
method may provide a useful test of approximate theories.
The major virtue of the density-matrix method is its relative
simplicity -- molecular simulations of entropic quantities
are difficult and, hence, rare.  Further calculations are
in progress.
\vspace{5mm}

It is a pleasure to thank E.G.D. Cohen, M.J. Feigenbaum,
N.N. Khuri and C.K. Zachos for several useful 
discussions, and Michael Wortis for some
particularly trenchant remarks.

\newpage
\begin{center}
{\Large{Table 1. Surface tension $\gamma$ (dynes/cm)}}
\end{center}

\vspace{3mm}
\begin{center}
\begin{tabular} {lllllll}

{\mbox{}} & {\mbox{}} & {\underline{DM}} & {\mbox{}} & 
{\underline{Experiment}} & {\mbox{}} & {\underline{SPT}}\\
{\mbox{}} &  {\mbox{}} & {\mbox{}} & {\mbox{}} & {\mbox{}} & {\mbox{}}
& {\mbox{}}\\ 
Sodium & {\mbox{}} & 164 (373$^0$K) & {\mbox{}} & 206 (371$^0$K) & {\mbox{}} 
& 91 (371$^0$K)\\
{\mbox{}} & {\mbox{}} & {\mbox{}} & {\mbox{}} & {\mbox{}} & {\mbox{}}
& {\mbox{}}\\
Potassium & {\mbox{}} & 93 (338$^0$K) & {\mbox{}} & 88 (337$^0$K) & {\mbox{}}
& 56 (337$^0$K)\\
{\mbox{}} & {\mbox{}} & {\mbox{}} & {\mbox{}} & {\mbox{}} & {\mbox{}}
& {\mbox{}}\\
Water & {\mbox{}} & 82 (298$^0$K) & {\mbox{}} & 72 (298$^0$K) & {\mbox{}} 
& 52 (303$^0$K)\\
{\mbox{}} & {\mbox{}} & {\mbox{}} & {\mbox{}} & {\mbox{}} & {\mbox{}}
& {\mbox{}}\\
Chlorine & {\mbox{}} & 31 (298$^0$K) & {\mbox{}} & 18 (293$^0$K) & {\mbox{}}
& 20 (293$^0$K)\\
{\mbox{}} & {\mbox{}} & {\mbox{}} & {\mbox{}} & {\mbox{}} & {\mbox{}}
& {\mbox{}}\\
Methane & {\mbox{}} & 21 (96$^0$K) & {\mbox{}} & 16 (95$^0$K) & {\mbox{}}
& -----------\\
{\mbox{}} & {\mbox{}} & {\mbox{}} & {\mbox{}} & {\mbox{}} & {\mbox{}}
& {\mbox{}}\\
Nitrogen & {\mbox{}} & 6.4 (77$^0$K) & {\mbox{}} & 9 (77$^0$K) & {\mbox{}}
& 9 (77$^0$K)\\

\end{tabular}
\end{center}

\newpage
\begin{center}
References
\end{center}

\begin{enumerate}

\item  J.D. Bekenstein, Phys. Rev. D{\bf 7}, 2333 (1973); {\bf 9},
3292 (1974); S.W. Hawking, Commun. Math. Phys. {\bf 43}, 199
(1975); G.W. Gibbons and S.W. Hawking, Phys. Rev. D{\bf 15}, 2752
(1977). 

\item  L. Bombelli, R.K. Koul, J. Lee, and R.D. Sorkin, Phys.
Rev. D{\bf 34}, 373 (1986).

\item  M. Srednicki, Phys. Rev. Lett. {\bf 71}, 666 (1993).
See also C. Callan and F. Wilczek, Phys. Lett.
B{\bf 333}, 55 (1994); V. P. Frolov, Phys. Rev. Lett. {\bf 74},
3319, (1995).

\item  J.S. Rowlinson and F.L. Swinton, {\em Liquids and
Liquid Mixtures} (Butterworth Scientific, London, 1982); J.P.
Hansen and I.R. McDonald, {\em Theory of Simple Liquids}
(Academic, New York, 1976).

\item  J.S. Rowlinson and B. Widom,
{\em Molecular Theory of Capillarity} (Clarendon Press,
Oxford, 1982).  See also R. Evans, Adv. Physics {\bf 28}, 143 (1979).

\item  L.S. Ornstein and F. Zernike, Proc. Akad. Sci. (Amsterdam)
{\bf 17}, 793 (1914).

\item D.J.E. Callaway and D.J. Maloof, Phys. Rev. D{\bf 27},
406 (1983);
D.J.E. Callaway, Phys. Rev. D{\bf 27}, 2974 (1983). 

\item R. C. Tolman, {\em The Principles of Statistical
Mechanics}, (Oxford, London, 1938), chapter 14.

\item R. Evans, et. al., Mol. Phys. {\bf 50}, 993 (1983).

\item  A.H. Narten and H.A. Levy, J. Chem. Phys. {\bf 55}, 2263
(1971),

\item  J.J. Jasper, J. Phys. Chem. Ref. Data {\bf 1}, 841 (1972).

\item  C.A. Croxton, {\em Introduction to Liquid State Physics},
(Wiley, New York, 1975).

\item  H. Reiss, in {\em Advances in Chemical Physics}, V. IX, ed.
I. Prigogine (Interscience, New York, 1965).

\item  J.P.M. Postma, J.H.C. Berendsen, and J.R. Haak, Faraday
Symp. Chem. Soc. {\bf 17}, 55 (1982).

\item  A.J. Greenfield, J. Wellendorf, and N. Wiser, Phys. Rev.
{\bf A4}, 1607 (1971).

\item  P.W. Schmidt and C.W. Tompson, in {\em Simple Dense Fluids},
ed. H. Frisch (Academic, New York, 1968).

\item  F.P. Buff and R.A. Lovett, in ref. [16].

\item  D.J.E. Callaway, PROTEINS: Structure, Function, and
Genetics {\bf 20}, 124 (1994).

\end{enumerate}
\newpage
\noindent
{\underline{Figure Caption}}

\vspace{.50in}
\noindent
{\underline{Figure 1}}

\vspace{5mm}
\noindent
Plot of $S_{out}(R)$ versus $R^2(\AA)^2$ for water.  Dashed line
is the function $2.50R^2$, displayed for comparison with plotted
data points.

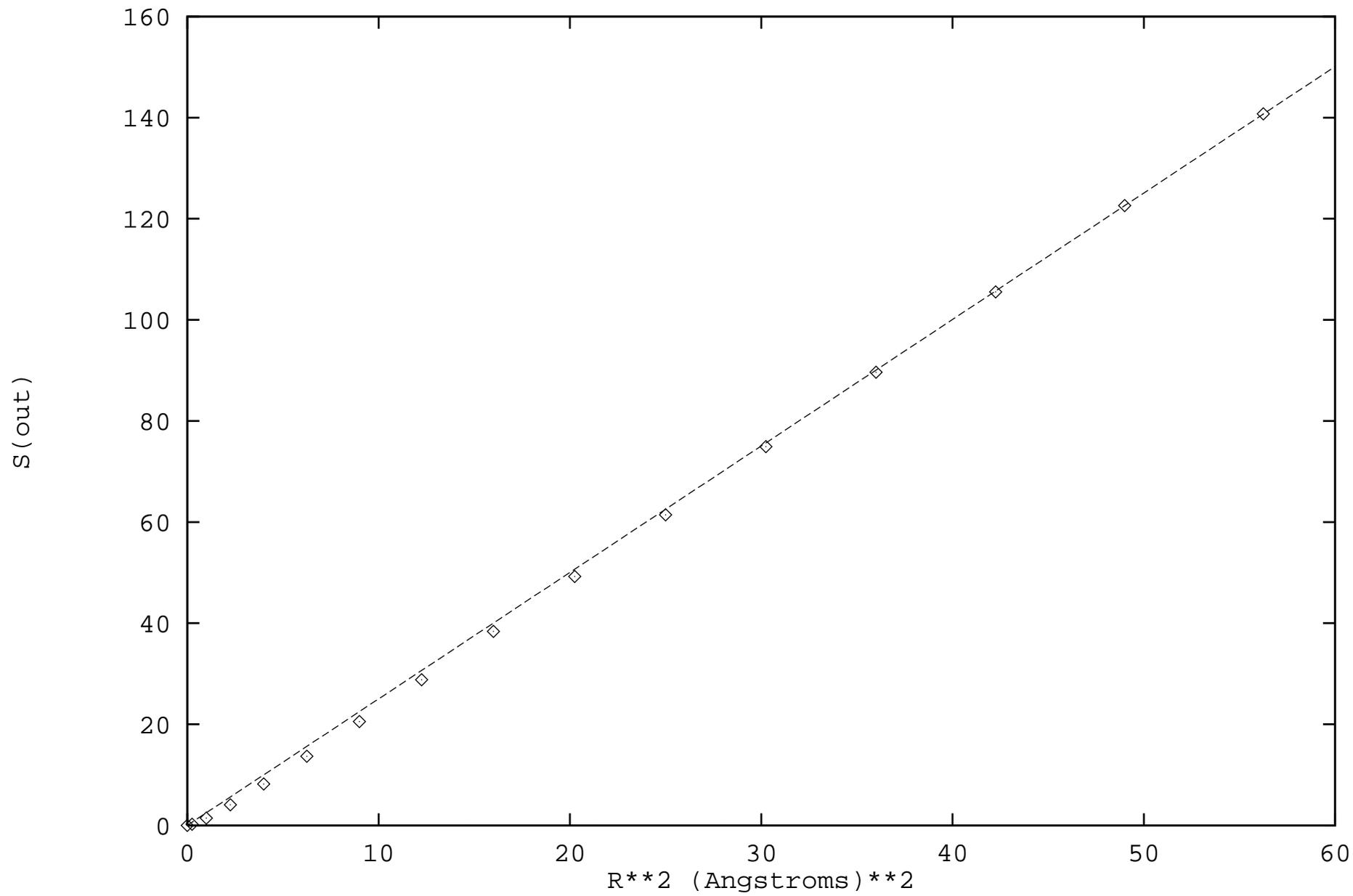
\end{document}